# Multi-objective Optimization of Cognitive Radio Networks

Rodney Martinez Alonso[1,2], David Plets[1], Margot Deruyck[1], Luc Martens[1], Glauco Guillen Nieto[2], Wout Joseph[1]

[1] Information Technology, Ghent University, Technologiepark Zwijnaarde 126, Ghent 9052, Belgium.
[2] R&D, LACETEL, Rancho Boyeros Ave. 34515, Havana 19200, Cuba.

Correspondence should be addressed to Rodney Martinez Alonso; rodney.martinezalonso@ugent.be

## Abstract

New generation networks, based on Cognitive Radio technology, allow dynamic allocation of the spectrum, alleviating spectrum scarcity. These networks also have a resilient potential for dynamic operation for energy saving. In this paper, we present a novel wireless network optimization algorithm for cognitive radio networks based on a cloud sharing-decision mechanism. Three Key Performance Indicators (KPIs) were optimized: spectrum usage, power consumption, and exposure of human beings. For a realistic suburban scenario in Ghent city, Belgium, we determine the optimality among the KPIs. Compared to a traditional Cognitive Radio network design, our optimization algorithm for the cloud-based architecture reduced the network power consumption by 27.5%, the average global exposure by 34.3%, and spectrum usage by 34.5% at the same time. Even for the worst optimization case, our solution performs better than the traditional architecture by 4.8% in terms of network power consumption, 7.3% in terms of spectrum usage and 4.3% in terms of global exposure.

## 1. Introduction

The lack of spectrum availability for satisfying the exponential increase in wireless traffic demand has become a major concern in the wireless communication community. Paradoxically, several extensive spectrum usage measurement campaigns have demonstrated that most of the radiofrequency spectrum is not in use or is sub-utilized. According to these spectrum surveys, performed in cities worldwide, the average use of the sub-3-GHz spectrum regarding both space and time is rarely higher than 20% [1, 2, 3, 4].

In this context, Cognitive Radio has become a flexible solution to overcome spectrum unavailability by opportunistically exploiting underutilized or unutilized spectrum [5, 6]. Research efforts on cognitive radio technologies have been undertaken to make use of the television broadcast *white spaces*. The TV *white spaces* (TVWS) are spectrum channels that are not in use by a primary licensed service (generally TV broadcasting) at a certain location during a certain period [7]. IEEE 802.11af [8], IEEE 802.22 [9] and IEEE 802.22b [10] standardize the secondary devices' cognitive radio features and technological requirements for dynamically accessing these TVWSs. These devices are not granted the licensed use of the spectrum, but regulatory authorities allow opportunistic spectrum access provided that they admit any interference (from primary licensed services) and do not cause harmful interference to the primary services themselves [11]. As such, interference minimization among these devices is a performance indicator for targeting when deploying cognitive radio networks.

Also for Long-Term Evolution (LTE) technologies, cognitive radio has been proposed as a solution towards more efficient use of the radioelectric spectrum. For instance, LTE-U allows the shared use of frequencies in the unlicensed ISM band as a way to overcome congestion issues in the 4G mobile networks and the lack of available spectrum [12]. The European Telecommunications Standard Institute (ETSI) technical report TR 103 067/2013 [13] analyzes the feasibility of LTE Cognitive Radio Systems operating in UHF band TVWS. Coexistence between LTE and TV broadcast systems is investigated as a typical use case of LTE with the cognitive ability for opportunistically using the UHF TV band [13].

Because of "spectrum scarcity", the bandwidth requirements for next-generation radio and 5G networks are only fully satisfied at the highest spectrum bands. As a consequence, several applications for 5G networks are limited in rural scenarios, small suburban cities, and in general in areas with low population density. The feasibility of 5G networks based on Cognitive Radio technology and carrier aggregation, enabling wideband access in rural areas has also been investigated. Hence, Cognitive Radio is a key enabling technology for new generation radio [14].

A major concern with Cognitive Radio is that interference management has been a challenge for scenarios with dense spectrum occupancy or high user density [15, 11, 7]. The currently implemented standards have some limitations for assessing the trade-off between interference and spectrum efficiency. For instance, IEEE 802.11f does not specify any spectrum sensing requirements to be met by user devices and base stations (BSs) for further spectrum allocation [8]. Although geolocation databases allow minimizing interference to the primary licensed services (generally digital television broadcasting), spectrum sensing, and spectrum allocation management have to be implemented to avoid interference to or from other Cognitive Radio devices in the same network or other Cognitive Radio networks. These databases are based on measurement campaigns and spectrum usage surveys. Hence, the spectrum usage information is not updated dynamically, reducing the potential spectrum sharing efficiency of cognitive radio networks. IEEE 802.22b provides some guidelines and mandatory channel sensing requirements. For instance, it defines the minimum requirements for the scheduling of sensing windows and quiet periods, mandatory users' device reports of detected interference and maximum interference thresholds for different signal types [10]. However, mandatory sensing techniques are not stipulated [16].

Several advances on spectrum sensing techniques for avoiding interference have been reported in recent years. However, no major advances have been reported on architecture and dynamic network optimization, being a limitation for a better trade-off among interference management and spectrum usage efficiency. Spectrum sensing techniques for cognitive radio applications are usually classified in blind (i.e., not taking into consideration the signal characteristics) and signal-specific sensing techniques [16]. The most basic spectrum sensing technique for cognitive radio is Energy Detection. However, the energy detection accuracy depends on prior knowledge of the noise level. The noise level at the detector is affected by several factors, i.e., temperature variation, calibration error, and low-noise amplifier gain variations [17]. Signal-specific sensing techniques are generally based on the correlation of the received baseband signal with a reference pattern [16]. This means that partial demodulation of the signal is required. Hence the hardware complexity and required sensing time increase.

In [18], the authors presented a cooperative scheme for spectrum sharing based on the information provided by secondary Wi-Fi nodes. Different improvements for increasing detection efficiency, reducing interference and exploration time are presented in [19, 20, 21]. A learning algorithm to improve spectrum exploration and to reduce the interference caused by cognitive radio devices is presented in [22]. [23] proposes a learning algorithm to maximize the network throughput by allowing varying sensing time and considering the historical behavior of the user's devices.

Few advances have been made on architecture and network medium access and connectivity efficiency to optimize the network's Key Performance Indicators (KPIs). In [24], authors present a novel cooperative system for the efficient usage of TVWS based on an Internet of Things (IoT) architecture. The proposed multilayer architecture improves coexistence issues and the protection of the primary services by combining spectrum sensing and a QoS feedback procedure implemented through a control logic in the IoT *social* platform (where all devices share performance information) [24]. This collaborative approach between the primary and secondary service could be an interesting solution to improve spectrum usage efficiency and coexistence among different services sharing the same spectrum bands (e.g., VHF/UHF bands).

Besides spectrum usage efficiency, power consumption and exposure of human beings to radiofrequency radiation are important network KPIs. These indicators are closely related to the environmental footprint of Information and Communication Technologies (ICT) [25, 26]. Hence, to achieve environment-friendly wireless networks, it is also required to optimize power consumption and exposure. However, these parameters require the assessment of a trade-off [27]. [28] presents a method for the efficient identification of multi-objective optimal settings on a wireless experimentation facility.

The novelty of this paper is the multi-objective optimization of new generation Cognitive Radio networks. Instead of a traditional distributed architecture for the spectrum management we consider a cloud-based architecture, allowing sharing the sensed information by all network devices. Based on the global knowledge of all devices it is possible to dynamically optimize the network, achieving a higher network efficiency in terms of power consumption, spectrum usage, and global exposure. The dynamic optimization of the network is required for improving its KPIs and reducing harmful interference from/to the primary licensed service. No research has been done yet, according to the authors' knowledge, on networking optimization to account for the trade-off among power consumption, spectrum usage efficiency and exposure for Cognitive Radio.

The outline of this paper is as follows. In Section 2 we describe the proposed cloud-based architecture for managing Cognitive Radio networks, briefly introduce Pareto Optimality, define metrics, details of rationale, describe the multi-objective optimization algorithm and a description of a realistic scenario and initial wireless network setup considerations. In Section 3, we present the network optimization results based on the proposed architecture and algorithm and benchmark the result against a traditional Cognitive Radio network. Conclusions are presented in Section 4.

## 2. Method

**2.1 Cloud-based architecture for Cognitive Radio network management**

Fig. 1 shows the proposed cloud-based architecture for the Cognitive Radio network that will allow achieving a higher multi-objective optimization level (Fig.1b) compared to the traditional architecture (Fig.1a).

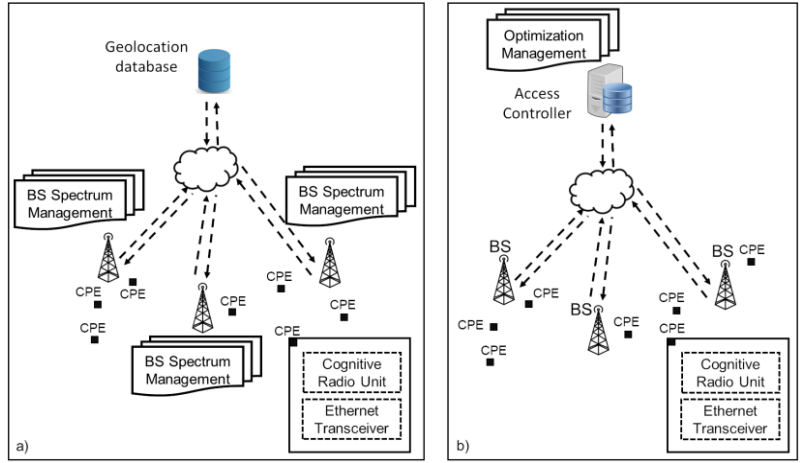

Fig. 1. Cognitive Radio network architecture a) Traditional b) Cloud-based management.

The traditional Cognitive Radio network architecture (in Fig.1a) comprises at the user side the Consumer Premise Equipment (CPE) including an ethernet transceiver and the Cognitive Radio unit. The architecture also includes the BSs including the BS management or system processing unit for handling the users' data registration, tracking, and application of spectrum management policies defined in the standard. Finally, a connection to ethernet or optical fiber provides access to the internet and (if applicable) to a geolocation database. In this topology (Fig. 1a), each BS oversees users' connections, registration, and tracking and the application of spectrum management policies. Each BS takes independent decisions about spectrum allocation based on the sensed data provided by users in their range and self-sensing. The provision of a geolocation database restricts the degrees of freedom of the BS, forbidding access to certain channels, footprint limitations, and other regulatory policies. This database is based on static data from spectrum surveys and does not provide real-time updates. The variations on the propagation conditions produce a variation on perceived network interference and spectrum usage efficiency.

Our architecture modification proposal for new generation Cognitive Radio networks consists of moving most management functions (mainly those related to spectrum management) to a central *Access Controller* (Fig.1b). In this way, it is possible to collect the information sensed by all network devices. By knowing and processing the network performance parameters from the whole network, it is possible to make better decisions (optimization) for serving the users' traffic with higher efficiency regarding the network power consumption, global exposure, and spectrum usage [29]. This information is sent by all BSs to the central *Access Controller* by the BS data backbone to the cloud. We assume the signalization data related to the Cognitive Radio functionalities is negligible compared to the backbone data capacity. The collected information is the same provided by the geolocation databases to the BSs. Hence, an adaptation of the headers and interfaces defined by the standards is not required.

Although a distributed network topology is generally efficient, if the devices taking decisions are affected by devices outside their control the system becomes unstable. This is similar to the concept of automated vehicular driving in 5G networks. When the network topology is based on a Vehicular-to-Vehicular (V2V) communication, each vehicle only has information about the vehicles in its range and the system quickly becomes unstable (cars start a crashing chain). This is because each car's control decision is based on the individual knowledge of the environment [29]. The same happens in the traditional Cognitive Radio network, where a crash corresponds to interference, and the same happens in any open-loop control system.

## 2.2 Pareto Efficiency

Pareto optimality is a concept of efficacy initially applied in social science and economic problems. The Pareto optimal state is defined as a state where it is not possible to make a single objective (parameter) better without making at least another one worse [30]. In engineering, usually more than one parameter needs to be maximized (multi-objective optimization problems). For a set of choices and a metric to value them, it is possible to find a set that is Pareto efficient. This set is named the Pareto Front [28]. Hence, it is possible to find a set of optimal trade-offs among all parameters depending on the design constraints, scenario, and application.

Wireless networks have several opposing performance indicators, e.g., throughput, energy, latency, electromagnetic radiation and spectrum usage [28]. The maximization of a certain parameter leads to the minimization of at least one other. This is generally a condition that is not optimal in many wireless applications. By using Pareto optimality, it is possible to evaluate several combinations of performance indicators, each one with a certain weight (Pareto coefficient) in the optimization algorithm (see Section 2.5). In this way, the general Pareto equation $P$ can be defined as a set of $n$ independent metrics $g$ multiplied by a certain weight $w$.

$$P(w_1; w_2; ...; w_n) = \{w_1 \cdot g_1; w_2 \cdot g_2; ...; w_n \cdot g_n\} \quad (1)$$

where for any combination of $w_1; w_2; ...; w_n$ the following condition must be satisfied:

$$\sum_{i=1}^{n} w_i = 1 \quad (2)$$

For having a continuous Pareto front and because Pareto coefficients can take an infinite number of values, we interpolate the solution set using Delaunay Triangulation. We investigate the tradeoff among three Key Performance Indicators (KPIs) in a cognitive radio network based on the IEEE 802.22 standard by means of Pareto Optimality, being the network power consumption, human exposure, and spectrum usage.

## 2.3 Rationale

In a cognitive network, each device has to sense the spectrum and provide to nearby BSs information related to perceived interference. This information is sent over a wireless link using BPSK modulation in a self-sensed empty channel to the nearby BSs. The first BS processing the user connection request and allocating the spectrum resources will register the user. In our proposal, this information is assumed to be collected by the BSs but the allocation of spectrum resources and interference management will be handled by a central access controller. Fig. 2 describes the process of connecting a single user to a Cognitive Radio BS.

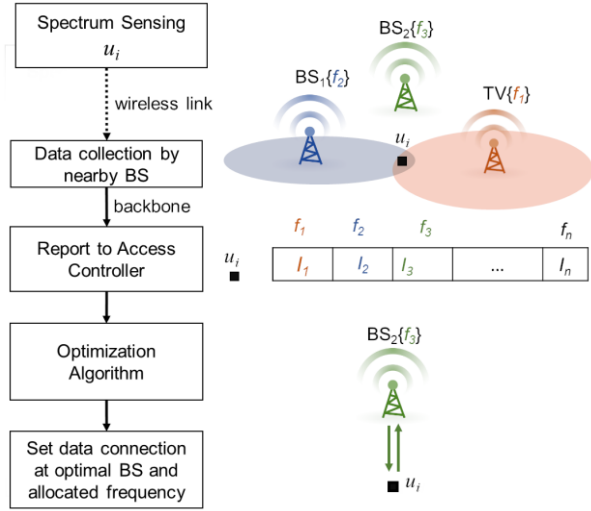

Fig. 2. Flow chart of the steps for connecting a user to a BS in the cloud-based Cognitive Radio architecture.

First, the user devices $u_i$ will sense the spectrum and will provide the perceived interference from other BSs, other users and broadcasting stations to a central access controller (i.e., $I_1$, $I_2$, $I_3$,…, $I_n$ in Fig. 2). In addition, the user provides information about the radiated signal, geolocation information and other parameters required by the standard [9] or by regulatory authorities. Notice that the initial link is settled with a nearby BS by using any frequency perceived by $u_i$ as free and BPSK modulation for minimum interference [9]. Further, the algorithm must assess the information received from all users and BSs, to compare the perceived interference with the maximum interference signal level allowed (ISL [dBm]) and settle the best connection for $u_i$ taking into account the network KPIs. The ISL is the maximum signal level for which the algorithm will allow reusing the same frequencies (i.e., $f_1$, $f_2$, $f_3$,…, $f_n$, in Fig. 2) by different BSs. One ISL constraint is defined for sharing frequencies used by television broadcasting (e.g., *TV* in Fig. 2) and another value for sharing frequencies in use by Cognitive Radio BSs (e.g., $BS_1$ and $BS_2$ in Fig. 2). Notice that the spectrum allocation is settled for each user. A certain BS can use different frequencies for communicating with different users at the same time, depending on the ISL constraint. For instance, if the interference is higher than the maximum allowable $BS_2$ and $u_i$ (see Fig. 2) must communicate in a frequency not in use by $BS_1$ or *TV* (i.e., $f_3$). Otherwise, it can reuse the same frequency (i.e., $f_1$ or $f_2$).

For the greenfield network planning, a reduction in the number of BSs has a major impact on the minimization of the network power consumption and cost. This is because the Cognitive Radio BSs have an *idle* or fixed power consumption that is not directly related to the radiated power or with the traffic load. For instance, for peak traffic and maximum radiated signal level, the Cognitive Radio BS power consumption is 64W versus 38W without data traffic (*idle* power consumption is approximately 59%) [31]. Additionally, the architecture with a centralized access control allows dynamically switching a group of BSs from *active* to *sleep* mode. The Cognitive Radio BS has a power consumption of only 9 W in *sleep* mode. However, by reducing the number of BSs, the radiated power per BS increases to cover the most distant users. As a consequence, the exposure in that cell increases [27]. In this way, the optimization algorithm should follow different optimization strategies depending on the parameters to be optimized. More details of the optimization algorithm are provided in Section 2.5).

**2.4 Metrics**

First, we defined the following metrics for accounting in the algorithm for the KPIs intended to be optimized, i.e., network power consumption, global exposure and spectrum usage).

The power consumption PC for a network configuration is accounted for following the power consumption model presented in [31] for a Cognitive Radio BS. Here, when accounting for the network power consumption, we consider that the centralized access controller (see Fig. 1) can switch the BSs that are not in use at a certain instant of time to *sleep* mode. In *sleep* mode, the Cognitive Radio BS has a power consumption as low as 9 W, including the radio unit, optical backhaul, and electrical transceiver. Notice that without a centralized controlling, not all power consuming components can be switched to *sleep* mode.

The global exposure $E_G$ is defined as a weighted average of the mean electric field $E_{50}$ and the 95-percentile of the field strength $E_{95}$ over the covered area [27], in order to optimize median and maximal exposure values. As in [25], we consider an equally weighted $E_{50}$ and $E_{95}$. Hence, the $E_G$ can be described by the following equation:

$$E_G = \frac{E_{50} + E_{95}}{2} \quad (3)$$

For calculating the electric field strength over the covered area, a grid of "test points" separated 50 m from each other is generated covering the entire map. At each grid point, the contribution of each transmitter $Tx_j$ to the electric field strength is calculated. The electric field $E_{Tx}$ [V/m] due to transmitter $Tx_j$ can be calculated based on $Tx_j$'s Equivalent Isotropically Radiated Power *EIRP* [dBm], frequency $f$ [MHz] and the path loss *PL* [dB] from the $Tx_j$ to the grid point [27], as described by Equation 4.

$$E_{Tx_j}(x,y) = 10^{\left(\frac{EIRP_{Tx_j} - 43.15 + 20 \cdot \log_{10}(f) - PL_{Tx_j}(x,y)}{20}\right)} \quad (4)$$

The contribution of all transmitters to the electric field strength at each grid point (*x,y*) is calculated by accounting the root sum of the squares of the electric field strengths due to each $Tx_j$ [27].

$$E_{total}(x,y) = \sqrt{\sum_{j=1}^{n}\left[E_{Tx_j}(x,y)\right]^2} \quad (5)$$

where *n* is the total number of transmitters. The field vectors caused by each source are thus assumed to have no phase correlation [27].

We define the spectrum usage $S_U$ as the number of channels that are required by the network for settling all simultaneous connections in the deployment scenario.

$$S_U(u_i, BS_j) = \sum_{ch=1}^{Smax} k_{ch}(u_i, BS_j) \tag{6}$$

where $k_{ch}$ is 1 if the channel *ch* has been assigned to a communication link between at least one user $u_i$ and a $BS_j$, else $k_{ch} = 0$. $S_{max}$ is the maximum number of channels allowed to be used according to the regulatory domain. The spectrum optimization is performed by reusing channels in the communication between each user and the BS. The metric $S_U$ is a measure of the spectrum occupation at the BS and user locations. Each user device dynamically accesses the spectrum at the most suitable frequency. A single BS can communicate with its connected users by means of different frequencies. The frequency channels are reused when the interference constraints (ISL) defined in the standard [9] are accomplished.

For a fair comparison among different solutions, we also define the white space availability in the whole area.

$$Wa(x, y) = S_{max} - S_{UTx}(x, y) \tag{7}$$

The white space availability *Wa* represents the number of channels available at each grid point (x,y) after the access controller assigned the spectrum for all links among the BSs and the users and also accounting for television stations. For calculating this value the whole area is divided into grid points with coordinates *x,y* (considering a resolution of 50m). *Wa* is the difference between the total number of channels $S_{max}$ and the number of channels in use by the Cognitive Radio network devices and the channels in use by the television broadcasting service $S_{UTx}$ at each grid point. Notice that $S_U$ accounts only for the Cognitive Radio network spectrum usage and $S_{UTx}$ also includes the television broadcasting spectrum usage. Hence, *Wa* represents the remaining channel availability in the area. The mean white space availability can be calculated accounting for *Wa* in all grid points.

**2.5 Multi-objective Optimization Algorithm**

Algorithm 1 describes the network optimization algorithm that will be used for minimizing the network power consumption, spectrum usage and exposure (i.e., goal KPIs). The algorithm is heuristic and capacity-based [25]. Hence, we cannot a guaranteed absolute optimal network solution but a solution that is good enough for solving the optimization problem. The solution convergence is defined by a maximum 2% standard deviation of the progressive average for each optimized network KPIs (spectrum, power consumption, and exposure). Notice that interference and spectrum management is not included in the base algorithm described in [25]. Hence, modifications were performed for the multi-objective optimization goals for this work.

The users and traffic requirements are input parameters. The algorithm first generates a uniform and pseudo-random distribution of the users in the area and assign a traffic load per user (see Phase 1 line 2 in Algorithm 1, the initial values for the input parameters are described in Section 2.6). Notice that the whole algorithm is repeated for a maximum number of simulations (*Max_Sim*). For the greenfield planning, an initial set of BSs (see initial values of input parameters in Section 2.6) is optimized to find the minimum required number of BSs ($N_{BS}$) [31] and the best BS locations (in terms of average path loss to users). A histogram with the number of connections settled by each BS during a group of simulations (*Max_Sim*) is performed, and the optimal BS locations are chosen based on the probability of having the lowest path loss to users (Phase 1 line 5 to 13, in Algorithm 1). The number of simulations *Max_Sim* is empirically chosen to guarantee that the progressive average of the KPIs has a standard deviation lower than 2%. The output of the first phase is a new set with a number $N_{BS}$ of optimal BS locations (see Phase 1 line 15 in Algorithm 1).

```
Phase 1:
Input: users, traffic, BSs, TV
1:     while Sim < Max_Sim
2:        Generate (users, traffic, BSs, TV)
3:     end while;
4:     while Sim < Max_Sim
5:        for u_i < users
6:           for BS_j < BSs
7:              PL = Path_Loss (u_i,BS_j)
8:              if (PL < Actual_Path_Loss) && Bitrate < Max_Bitrate
9:                 Best_BS (BS_j)
10:             end if;
11:          end for;
12:          Histogram [Best_BS]++;
13:       end for;
14:    end while;
Output:
15     Best_BS_locations ← get_Histogram(N_BS);

Phase 2:
Input: Best_BS_locations;
1:     Generate (Best_BS_locations);
2:     while Sim < Max_Sim
3:        for u_i < users
4:           for BS_j < BSs
5:              Calculate_Power_Consumption (u_i,BS_j);
6:              Calculate_Exposure (u_i,BS_j);
7:              Evaluate_ISL;
8:              Set_Spectrum_Allocation (u_i,BS_j);
9:              if fit (u_i,BS_j) > current_fit
10:                if Active (BS_j) && Bitrate (BS_j)
11:                   connect;
12:                end if;
13:             end if;
14:             if unconnected
15:                set_active BS (best_fit);
16:                connect;
17:                load_balance;
18:             end if;
19:          end for;
20:       end for;
21:       for u_i < users
22:          while connected (u_i)
23:             BS_Radiated_Power--;
24:          end while;
25:       end for;
26:       calculate_W_a (Network_Solution);
27:       generate (Network_Solution);
28:    end while;
Output
29:    Network_Solutions;
```

Algorithm 1. Multi-objective optimization algorithm.

The second phase receives as input the set of BSs (Best_BS_locations). The users and traffic densities are the same as Phase 1. In this phase, a dynamic optimization of the network is performed. For the total number of simulations *Max_Sim* and for each user ($u_i$), the algorithm calculates a fitness value for each possible connection $BS_j$. For each possible connection, the algorithm calculates the Power Consumption, Exposure, and after evaluating a certain Interference Signal Level constraint (ISL), it finds the best spectrum allocation for each link to be settled (Phase 2 lines 5 to 8 in Algorithm 1). For the network modeling and optimization, the interference levels are calculated based on the radiated power and path loss calculations. We start by evaluating (and further assigning) the lowest available frequency (for a better signal propagation). First, the algorithm will verify the interference level at the user and BS sites generated by all users, all BSs and television broadcasting towers in the surroundings. This is possible because we consider a centralized access architecture retrieving information from all devices. Hence, the decision on frequency assignment is based on the information provided by all BSs from all users. Notice that in the traditional Cognitive Radio network architecture, the BSs are in charge of the frequency allocations taking into consideration static information from a geolocation database and the interference level information provided only by users in their range.

The fitness function *fit* (Phase 2 line 9 in Algorithm 1) accounts for the network power consumption $P_C$ [W], global exposure $E_G$ [V/m] [25] and spectrum usage $S_U$ if the user $u_i$ is connected to a certain $BS_j$ (see metrics definition in Section 2.4).

$$fit(u_i; BS_j) = w_1 \cdot \left(1 - \frac{P_C}{P_{max}}\right) + w_2 \cdot \left(1 - \frac{E_G}{E_{max}}\right) + w_3 \cdot \left(1 - \frac{S_U}{S_{max}}\right) \quad (8)$$

where $P_{max}$ [W] is the maximum power consumed by the network (i.e., all BSs *active* with a maximum radiated power); $E_{max}$ [V/m] is the maximal exposure over the considered area for the same network conditions. In this way, all performance indicators are normalized, and no parameter is overrated. Hence, for the worst-case $P_C = P_{max}$, $E_G = E_{max}$, and $S_U = S_{max}$ the fitness function equals 0. The weight factors are defined as $w_1$, $w_2$, and $w_3$. These weight factors corresponding to the Pareto coefficients and range from 0 to 1. For the Pareto optimality, several combinations of weights must be evaluated. Hence, several simulations are performed depending on the resolution considered for the weight factors. We consider a resolution of 0.25 for the Pareto coefficients, yielding 15 possible fitness functions for each ($u_i$;$BS_j$) possible connection. Notice that in any coefficient combination the sum of $w_1$, $w_2$, and $w_3$ must be equal to 1.

Each user is connected to the BS with the highest fitness value (lowest power consumption, spectrum usage, and exposure), if this BS is already *active* and still can support the user's demanded bitrate (Phase 2 lines 9 to 13 in Algorithm 1). If no *active* BS can support the user's demanded bitrate, a new BS with the best fitness value is switched *active*. For balancing the network load, already connected users can be switched to this new *active* BS, if their fitness value to this BS is higher than before (Phase 2 lines 14 to 18 in Algorithm 1). Once all users have been evaluated, the first network solution is optimized by decreasing the BS radiated power (Phase 2 lines 21 to 25 in Algorithm 1). The stop condition is reached when the path loss experienced by a user is higher than the maximum allowable path loss [25]. The decrease of the radiated power will decrease the power consumption, exposure and will allow a better re-usage of the spectrum.

Finally the algorithm will calculate the network solutions white space availability (*Wa*) and will generate the network solutions (Phase 2 lines 26 and 27 in Algorithm 1). Notice that the density of users per area must be large enough to guarantee that the progressive average of the network KPIs converges after a reliable number of simulations with a low standard deviation (<2%).

## 2.6 Evaluation scenario and initial setup

To validate the proposed architecture and optimization algorithm, we modeled and optimized a Cognitive Radio network, in a real suburban wireless scenario. We consider the city of Ghent, Belgium (68 km$^2$) for the green field planning and later dynamic optimization of the network. Fig. 3b shows a map of Ghent City and the BS possible locations denoted with red squares. Fig. 3a shows a map of the Region of Flanders, Belgium covering an area of approximately 13,522 km$^2$ with the location of the television broadcast transmitters.

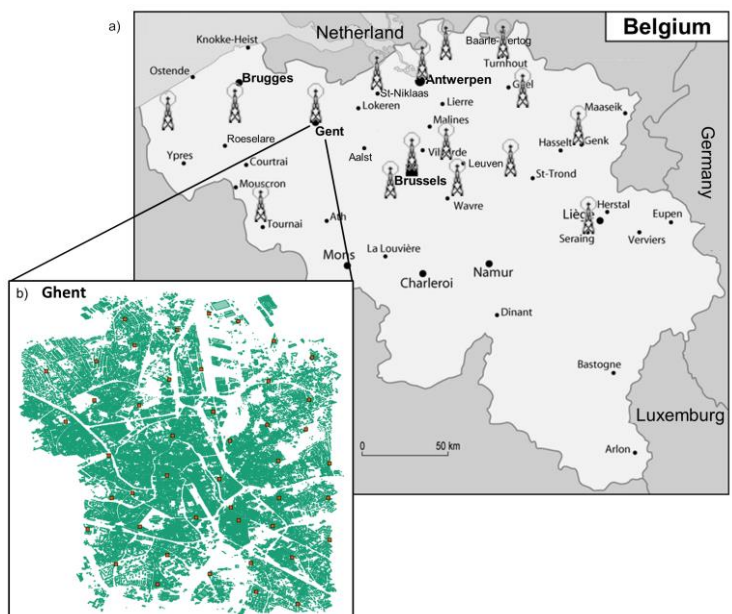

Fig. 3. a) Broadcast transmitters in the region of Flanders, Belgium and b) area to be covered in Ghent, City, 45 possible BS locations identified with red squares.

A traditional Cognitive Radio network design is also modeled for a comparison reference. For this network, it is not possible to implement the *sleep* mode. This is because in *sleep* mode the spectrum management and user tracking in the BS are switched off. Without a centralized access controller capable of assuming spectrum management and tracking functions the Cognitive Radio BS can only be switched to *idle*. In *idle* mode, the BSs implement most sensing functions, signaling and tracking, but no user data payload traffic is handled.

A set of 45 possible locations is considered for the Cognitive Radio BSs (represented by squares in Fig. 3b). After the first optimization step, the algorithm retrieves a histogram of the optimal BSs locations and determines the minimum number required to satisfy the intended coverage ($N_{BS}$, see Section 2.5). The network is designed to guarantee a 95% coverage at the cell-edge during 99% of the time. We consider the link budget presented in [31] for the IEEE 802.22 standard.

We consider 224 simultaneous user connections at the peak traffic time and a bit rate of 1 Mbps per user [31]. Three scenarios are modeled, and the networks are dynamically optimized for three densities of BS infrastructure availability, i.e., the minimum number of BSs that satisfies the coverage constraint ($N_{BS}$), 25%, and 50% higher BS density. In all cases, for the green field network planning the BSs are chosen according to the user's connections histogram.

The algorithm will realize two network solutions for each BS infrastructure availability considering an interference threshold of -116 dBm and -93 dBm for the Cognitive Radio sensed signals. The threshold of -116 dBm is considered if channel occupancy is based on sensing the IEEE 802.22.1 Beacon and sensing mode 0. The threshold of -93 dBm is defined in the standard for sensing mode zero and signal type IEEE 802.22 WRAN [9]. There is no recommended interference threshold value for DVB-T/T2 digital television signals in the IEEE 802.22 standard [9, 10]. For evaluating the re-usability of frequencies in use by nearby television broadcasting stations, an interference constraint of -95 dBm is considered. This value is based on the recommended protection contour of the broadcast transmitter [3]. The chosen interference constraint guarantees the minimum carrier-to-interference-ratio recommended in [11] for the protection of the primary service from harmful interference.

All the broadcast transmitters around Ghent are included in the model to account for the interference levels (Fig. 3a). For the path loss calculations for TV Towers in Flanders, we consider their actual transmitter configurations [32, 33] and the ITU path loss model ITU-R P.1546-5/2013 [34]. For the path loss calculations in Ghent, we consider an experimental one-slope path loss model based on extensive measurement campaign in the UHF band as described in [35]. This model has a higher precision for Ghent city than the ITU-R P.1546-5/2013 model.

## 3. Results and Discussion

The minimum number of BSs to satisfy the coverage (spatial and temporal) of the Cognitive Radio network is $N_{BS}$ = 22. The second optimization step is realized for the chosen 22 BSs based on the computed histogram (see Section 2.5). The dynamic network optimization from Phase 2 (in Section 2.5) is also realized for a 25% (28 BSs) and 50% (33 BSs) higher BS infrastructure (BS selection based on the histogram).

### 3.1 Pareto Optimality

Fig. 4 shows the generated objective optimization results for a) 22 BS (minimum to guarantee coverage requirements, $N_{BS}$), b) 28 BS (increase of BS locations by 25%) and c) 33 BS (increase of BS locations by 50%) with an Interference Signal Level constraint ISL = -93 dBm.

The best performance in terms of spectrum and global exposure is denoted by markers 2 and 3, respectively. The difference for a given BS density between both markers is maximally 5.3%. This is because a major impact in both spectrum usage and exposure is achieved for a larger density of *active* BSs with a low radiated power. For the same reason, by increasing the infrastructure availability (more BSs to be switched from *sleep* to *active*) by 25%, the spectrum usage is reduced by 3.7% (see marker 2 in Fig. 4a vs Fig. 4b) and the global exposure reduces by 16.3% (see marker 3 in Fig. 4a vs Fig. 4b). An increase of 50% of BS density (Fig. 4c) reduces the spectrum usage by 5.6% (see marker 2) and exposure by 15.8% (see marker 3) compared with 22 BS density (Fig. 4a). Notice that here the exposure is similar for 28 BS and 33 BS, in this case the reduction of radiated power per BS does not compensate for the increase of radiating sources. The improvements in spectrum usage and exposure have a drawback on the network power consumption. The power consumption increases by 13.3% to 15% for a 25% higher BS density (Fig. 4b marker 2 and 3) and 17.7% to 20.6% for a 50% higher BS density (Fig. 4c). This is a direct consequence of a higher density of *active* BSs for the network solutions in Fig. 4b and Fig. 4c.

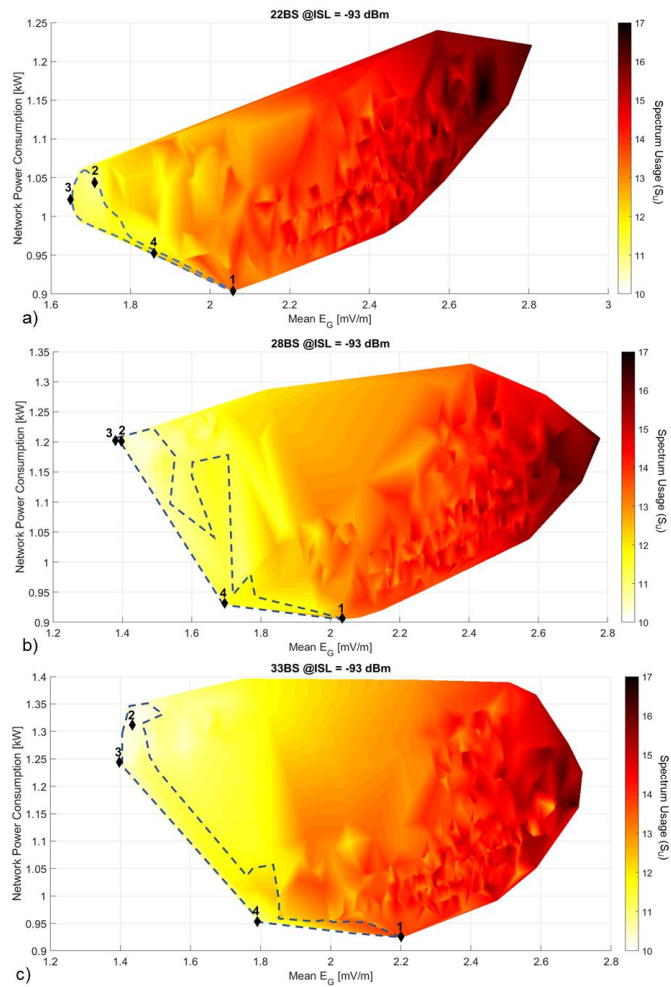

Fig. 4 Optimization results for a) 22 BS b) 28 BS and c) 33 BS locations @ ISL = -93 dBm (2D projection view). The dashed line represents (part) of the Pareto front. Marker 1 denotes the best power consumption results, marker 2 best spectrum usage result, marker 3 best exposure result and marker 4 optimal trade-off.

For the best network solution in terms of power consumption (see Fig. 4 marker 1) there is no significant variation on this KPI as the density of available BS locations increases. The network power consumption varies from 0.91 kW to 0.92 kW (the difference is lower than the standard deviation). This is because the algorithm switches most of the BSs to *sleep* mode and increases the radiated power to reach the farthest users. The lower density of *active* BS (serving user traffic) leads to a lower power consumption even when the radiated power increases. As the density of *active* BSs decreases, the power consumption is reduced from 30% to 35% compared with the maximum network power consumption, but the spectrum increases around 20% and the exposure from 20% to 37% (see marker 1 in Fig. 4).

The most balanced trade-off among the KPIs is denoted by marker 4. Increasing the number of BS locations do not allow a significant improvement of the mean point in the Pareto frontier (marker 4). Notice that for a 25% higher BS density (see Fig. 4b marker 4) the mean spectrum usage only improves by 3.3% and the global exposure by 8.6% while the network power consumption remains almost equal (the difference is lower than the standard deviation). For a 50% higher BS density (see Fig. 4c marker 4) the mean spectrum usage improves by 10% but the global exposure only improves by 3.7% with almost the same network power consumption.

In general, with the increase in the number of *active* BSs, spectrum and network exposure improve for different points in the Pareto front (surface inside the dashed line), but with the drawback of a higher power consumption. In fact, the optimization strategies for power consumption and exposure are contradicting. For reducing global exposure, the algorithm leads to a higher number of *active* BSs with low radiated power. For reducing power consumption, the algorithm leads to a lower number of *active* BS. This is because the *sleep* mode of the BS consumes only 9W (14% of the maximum power consumption). Hence, a higher optimization, in terms of power consumption, is achieved when BSs are switched to *sleep* mode. In addition, for improving spectrum reusability, a lower radiated signal level will lead to lower spectrum usage but with a more balanced rate than in the case of the network exposure optimization. This is because the ISL constraint has an additional impact on the connection decisions and spectrum allocation.

### 3.2 Cloud-based vs traditional network architecture

Fig. 5 shows the difference (in percentages) for each pareto point in the cloud-based architecture and the traditional distributed architecture for 22 BS.

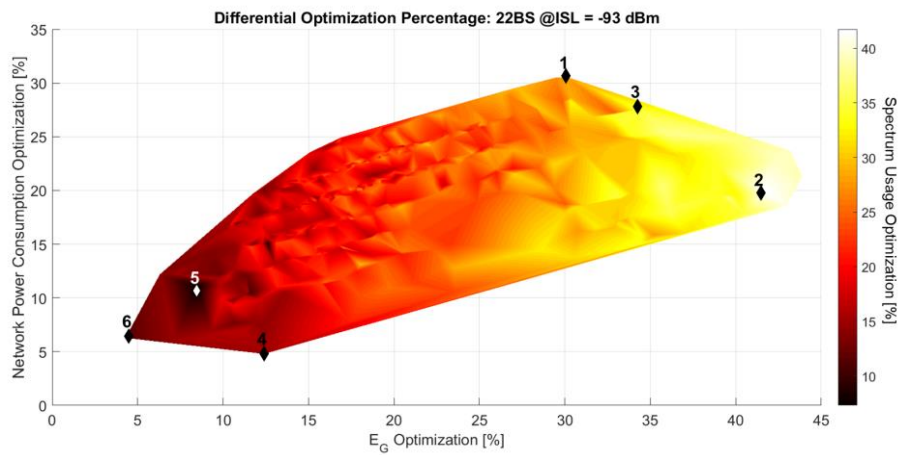

Fig. 5 Differential Optimization comparison between the traditional Cognitive Radio network and the results for the cloud-based Cognitive Radio architecture for 22 BS @ ISL = -93 dBm (2D projection view). Marker 1 denotes the best power consumption result, marker 2 the best spectrum usage result, marker 3 the best exposure result, marker 4 the worst-case power consumption result, marker 5 the worst-case spectrum usage result, and marker 6 the worst-case exposure result.

The network modeling for the traditional Cognitive Radio network and 22 BS, yielded a power consumption of 1.3 kW, a network global exposure of 2.9 mV/m and mean spectrum usage of 18.4 channels. In Fig. 5 we represent how high these values are compared with each Pareto point in the optimized cloud-based Cognitive Radio network. For maximum network power consumption saving, the cloud-based architecture reduces power consumption by 30.6% (see marker 1 in Fig. 5). For this condition the spectrum usage is 28.1% lower and the mean global exposure is 30.1% lower, compared to the traditional Cognitive Radio network architecture. The best trade-off among spectrum usage and mean global exposure (see marker 2 Fig. 5) achieves a higher performance by 41.6% (for both KPIs), and lower power consumption by 19.9%. The most balanced network solution among all KPIs (marker 3 in Fig. 5) achieves a lower spectrum usage by 34.5%, lower global exposure by 34.3% and lower network power consumption by 27.5%. In addition, notice that for the worst-case network solution for each KPI the proposed architecture performs better than the traditional Cognitive Radio network at least by 4.8% in terms of network power consumption (marker 4 in Fig. 5), 7.3% in terms of spectrum usage (marker 5 in Fig. 5), and 4.3 % in terms of global exposure (marker 6 in Fig. 5). These results are due to the fact that a better connection decision is made by a centralized access controller when assessing the data collected by all devices, rather than a local decision based only on information from devices in the BSs service area.

### 3.3 Effect of ISL constraint

Fig. 6 shows the Pareto multi-objective optimization results for 22 BS (minimum to guarantee coverage requirements) with an interference signal level constraint of -116 dBm (23 dB lower than before at -93dBm).

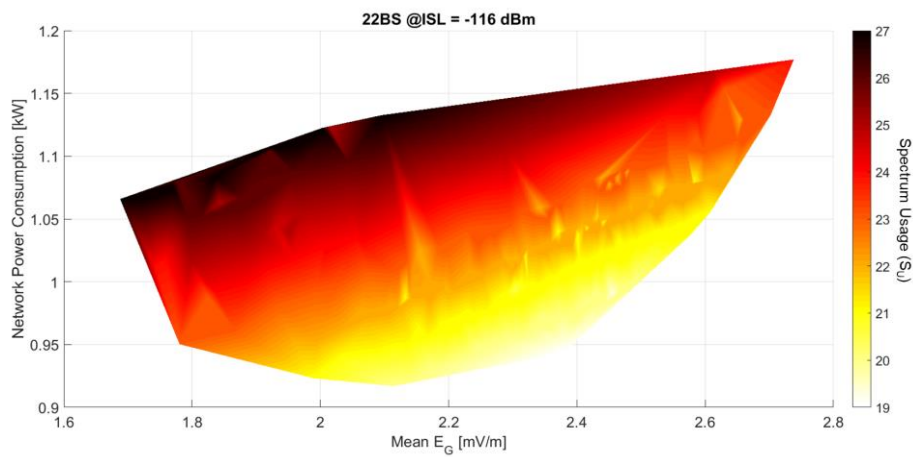

Fig. 6 Pareto optimization results for 22 BS @ ISL = -116 dBm

The optimization results achieved for the network power consumption and global network exposure are similar in comparison to the results for an interference constraint of -93 dBm (see results for ISL = -93 dBm in Fig. 4a). However, the spectrum usage is considerably higher. For the best network solution in terms of spectrum usage the $S_U$ is equal to 19.0. This value is 43.7% worse than the best spectrum usage results for 22 BSs and ISL = -93 dBm, and 10% worse compared to the worst case in terms of spectrum for 22 BS and ISL = -93 dBm. This is because an increase in the number of *active* BSs and a decrease in the radiated signal level per BS, is not enough to allow a high reuse of the spectrum, due to the 23 dB stricter ISL constraint.

### 3.4 White Space Availability

Fig. 7 shows the white space distribution map (white space availability at each grid point in the area, see metric on Section 2.4) for the traditional Cognitive Radio network with non-coordinated spectrum management and 22 BS (Fig. 7a), the cloud-based centralized spectrum management with 22 BS (Fig. 7b) and the same network architecture with 33 BS (Fig. 7c).

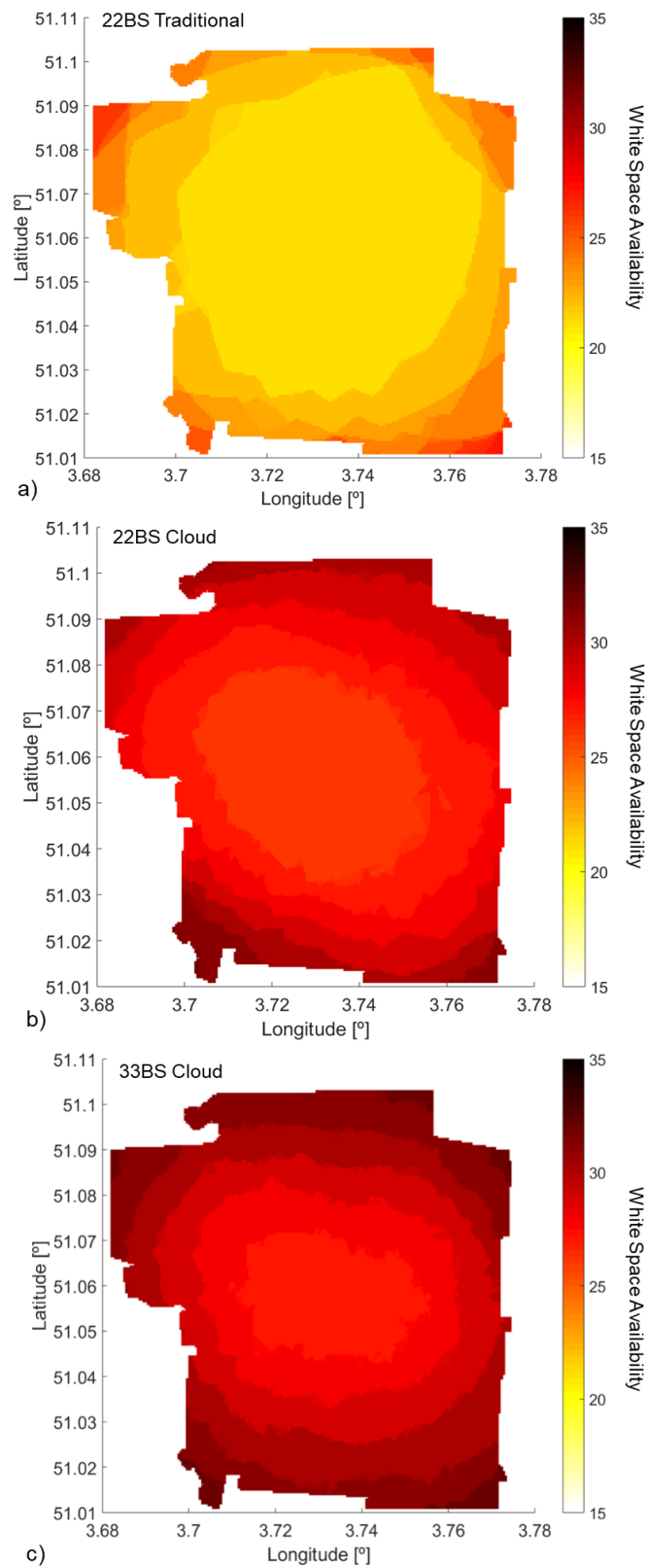

Fig. 7 White space availability distribution for a) the traditional Cognitive Radio architecture with 22 BSs, b) the proposed cloud architecture with 22 BSs and c) 33 BSs.

The white space distribution maps in Fig. 7b and Fig. 7c correspond with the best results in terms of spectrum usage (denoted with marker 2 in Fig. 4a and 4c, respectively). The mean white space availability (accounting for the mean in the whole area) is 26.3% higher for the optimized proposed architecture than in the traditional Cognitive Radio network (Fig. 7b and 8a respectively). Nevertheless, for the balanced optimality, the improvement is slightly lower (22.8%). Notice that in the city center the difference of white space availability can be higher. This is because the interference level is higher in the city center due to the BS locations distribution and the confluence of the radiation from more BSs.

Although the mean white space availability does not increase significantly with the number of BS infrastructure availability (approximately 2% for a 50% increase in BS availability), the gradient (speed) of white space availability is higher for 33 BS than for 22 BS (Fig. 7b and 7c, respectively). This means more channels will be available in a region closer to Ghent if 33 BS locations are used for the green field network planning instead of 22 BS. This is because, although there are more radiating sources (*active* BSs), the radiation level per BS is lower, hence the radiation is concentrated in a smaller area due to the environment path loss.

## 4. Conclusions

By means of a novel multi-objective optimization algorithm for Cognitive Radio networks, we quantified the advantages of cloud-based network management for Cognitive Radio technologies in comparison with a traditional distributed architecture. A Pareto efficiency modeling is performed for quantifying the trade-off among three KPIs: Power Consumption, Spectrum Usage, and Exposure.

Compared to a traditional Cognitive Radio network, our proposed architecture and optimization algorithm reduces the network power consumption by 27.5%, the average global exposure by 34.3% and spectrum usage by 34.5% for the best balance among the three KPIs. Important is to notice that even for the worst pareto point, our solution performs better than the traditional architecture by 4.8% in terms of network power consumption, 7.3% in terms of spectrum usage and 4.3% in terms of global exposure.

For the cloud-based architecture a higher BS infrastructure density (beyond the minimum that guarantees the intended spatial and temporal coverage) improves spectrum usage up to 5.6% and global exposure up to 16.3% but with a drawback in terms of network power consumption from 13.3% to 20.6%.

Future research will consist of the experimental characterization of the data rate as a function of the interference for a dynamic interference constraint assessment for Cognitive Radio networks.

## Conflicts of Interest

The authors declare that there is no conflict of interest regarding the publication of this paper.

## Funding Statement

R. Martinez Alonso is supported by LACETEL, and a doctoral grant from the Special Research Fund (BOF) of Ghent University, Belgium.